\documentclass[proceedings]{JHEP3} % 10pt is ignored!
\usepackage{epsfig}   % please use epsfig for figures
\def\beq{\begin{eqnarray}}    %%%  begequation/eqnarray
\def\eeq{\end{eqnarray}}      %%%  endequation/eqnarray

\def\be{\beta}

\def\de{\delta}

\def\La{\Lambda}
\def\la{\lambda}
\def\na{\nabla}
\def\pa{\partial}

\def\si{\sigma}

\def\Ga{\Gamma}

\def\La{\Lambda}

\conference{International Workshop on Astroparticle and 
High Energy Physics}

\title{The anomaly-induced effective action and natural inflation}

\author{
A.M. Pelinson
\thanks{Ins. de Astronomia, Geofísica e Ciências Atmosféricas, 
Universidade de S\~ao Paulo, S\~ao Paulo, Brazil}
, $\,$
\speaker{I.L. Shapiro}
\thanks{Dep. de F\'{\i}sica, ICE, 
Universidade Federal Juiz de Fora, Minas Gerais, Brazil}
, $\,$
J. Sol\`a 
\thanks{Dep. E.C.M. and CER for Astrophysics, Particle Physics 
and Cosmology, Universitat de Barcelona, Spain}, $\,$
, $\,$
F.I. Takakura
\thanks{Dep. de F\'{\i}sica, ICE, 
Universidade Federal Juiz de Fora, Minas Gerais, Brazil}}

\abstract{
The anomaly-induced inflation (modified Starobinsky model)
is based on the application of the effective quantum field 
theory approach to the Early Universe. We present a brief 
general review of the model and show that it does not 
require a fine-tuning for the parameters of the theory or 
initial data, gives a real chance to meet a graceful exit 
to the FRW phase and also has positive features with respect
to the metric perturbations.}

\begin{document} 
\providecommand{\href}[2]{#2}

%%%%%%%%%%%%%%%%%%%%%%%%%%%%%%%%%%%%%%%%%%%%%%%%%%%%%%%%%%
\section{Introduction}

The possibility of observing phenomena occuring in the 
early universe, and in particular during inflation 
\cite{Guth,linde} give a chance to learn new information 
about the high energy physics. One of the options is to 
consider the model of inflation which contains smaller 
phenomenological input compared to the conventional 
inflaton models (see, e.g. \cite{KT}) and can be directly 
deduces from the results of quantum field theory (QFT) in curved
space-time. The remarkable example of such model is based
on the vacuum quantum effects, in the 
simplest case on the effects of massless fields. In this
case the leading quantum phenomenon is conformal anomaly.  
%%%%%%%%%%%%%%%%%%%%%%%%%%%%%%%%%%%%%%%%%%%%%%%%%%%%%%%%%%%
The original version of the anomaly-induced inflation 
\cite{fhh,star,vile,ander} has been developed in 80-ties
is the cosmological model which 
takes into account the vacuum quantum effects of the free, 
massless and conformally coupled to metric matter fields
\cite{birdav}. The quantum correction to the Einstein 
equation
\beq
R_{\mu\nu}\,-\,\frac12\,R\,g_{\mu\nu}\,=\,
8\pi G\,<T_{\mu\nu}> \,-\,\La
\label{1}
\eeq
(where we added the cosmological constant (CC) 
for the sake of generality)
produces a non-trivial effect because the anomalous trace
of the stress tensor 
\beq
T\,\,=\,<T_\mu^\mu>\,=\,
- \,(wC^2 + bE + c{\nabla^2} R)\,,
\label{main equation}
\eeq
where
\beq
w \,=\, \frac{1}{(4\pi)^2}\,\Big(
\frac{N_0}{120} + \frac{N_{1/2}}{20} + \frac{N_1}{10} \Big)\,,
\nonumber
\\
b \,=\, -\,\frac{1}{(4\pi)^2}\,\Big( \frac{N_0}{360} 
+ \frac{11\,N_{1/2}}{360} + \frac{31\,N_1}{180}\Big)\,,
\nonumber
\\
c \,=\, \frac{1}{(4\pi)^2}\,\Big( \frac{N_0}{180} + \frac{N_{1/2}}{30}
- \frac{N_1}{10}\Big) \,.
\label{c}
\eeq
In the absence of matter, one can obtain the cosmological 
solution in two distinct ways: using the $\,(00)$-component 
\cite{fhh,star} or via the anomaly-induced effective 
action \cite{buodsh,book}. Indeed, the last option is 
completely equivalent to taking the trace of (\ref{1}). 
The resulting equation has, for $\,k=0$ FRW metric, 
the following form (since the cases $k=\pm 1$ are 
quite similar \cite{asta} we will not consider them 
here):
\beq  
\frac{{\stackrel{....}{a}}}{a}
+\frac{{3\stackrel{.}{a}} {\stackrel{...}{a}}}{a^2}
+\frac{{\stackrel{..}{a}}^{2}}{a^{2}}
-\left( 5+\frac{4b}{c}\right) 
\frac{{\stackrel{..}{a}} {\stackrel{.}{a}}^{2}}{a^3}
-\frac{M_{P}^{2}}{8\pi c}
\left( \frac{{\stackrel{..}{a}}}{a}+
\frac{{\stackrel{.}{a}}^{2}}{a^{2}}
-\frac{2\Lambda }{3}\right)\,=\,0\,,
\label{foe}
\eeq
where $M_P=G^{-1/2}$ is a Planck mass.
The equation above has remarkable particular solution
\beq
a(t) \,=\, a_0 \cdot \exp(Ht)
\label{flat solution}
\eeq
where (motivated by the recent supernova data \cite{SN}), 
we consider only positive CC in the low-energy regime
\beq
H\,=\, \frac{M_P}{\sqrt{-32\pi b}}\,\left(1\pm 
\sqrt{1+\frac{64\pi b}{3}\frac{\Lambda }{M_P^2}}\right)^{1/2}.
\label{H}
\eeq
As far as $\,\La \ll M_P^2$, we meet two very different 
solutions 
\beq
H_{c}\,=\,\sqrt{\frac{\Lambda }{3}}
\label{H now}
\\
\,\,\,\,\,\,\,\,\,\,\,\,{\rm and}\,\,\,\,\,\,\,\,\,\,\,\,
H_S\,=\,\frac{M_P}{\sqrt{-16\pi b}}\,.
\label{HH}
\eeq
The first solution (\ref{H now}) is exactly the one in the 
theory without quantum corrections, while the second solution 
$\,H_S\,$ is the inflationary solution of 
Starobinsky. We suppose that the first solution corresponds, 
approximately, to the present-day universe and the second 
one to the beginning of the inflation. Hence, our purpose
will be to find a natural interpolation between them.

Let us consider the initial phase of inflation, where 
the CC plays no much role. The equation for the $(00)$
component, equivalent to the Eq. (\ref{foe}), has been
completely studied by Starobinsky \cite{star}. The phase 
portrait of the theory may look quite different depending 
on the sign of the coefficient $\,c$ \cite{asta}. The 
inflationary solution with $H_S$ is stable for $\,c>0\,$ 
and unstable for $\,c<0$. The phase diagram for $\,c>0\,$
is presented at the Fig. 1. It is easy to see that there
is only one (inflationary) attractor, therefore stable
inflation does not depend on the initial conditions
(except the need to start from the homogeneous and 
isotropic metric). 

\FIGURE[t]{\epsfig{file=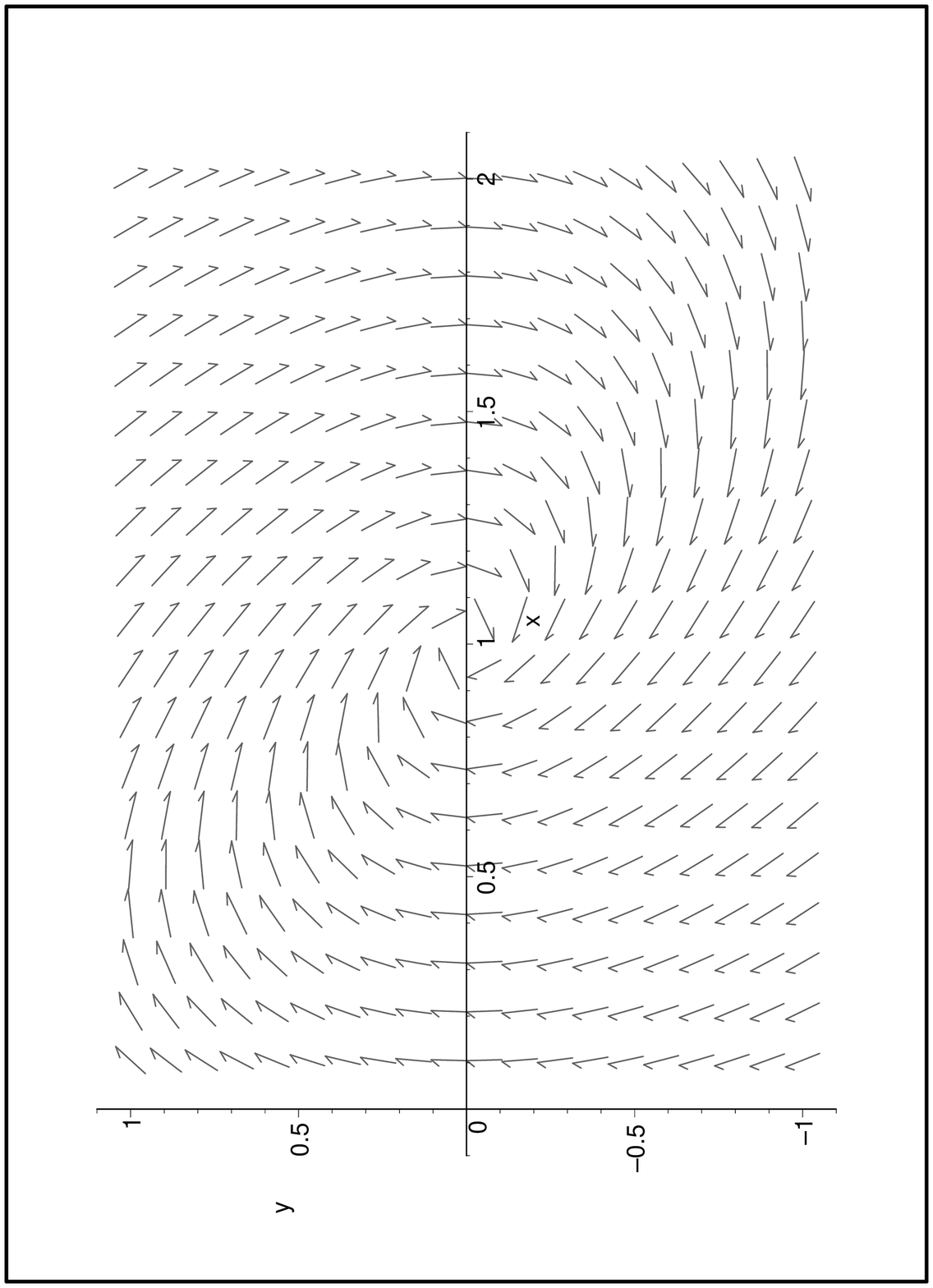, width=.4\textwidth}
\caption{\sl The phase diagram for the 
stable version of the Starobinsky model.}
\label{Figure 1}}	

In the unstable case the
phase diagram can be found in \cite{star}. There are 
several distinct attractors, one of which corresponds 
to the FRW evolution \cite{star} and others to the 
unphysical, run-away type solutions \cite{wave}. 
The original Starobinsky model deals only with the 
unstable solution and implies that the initial conditions 
must be chosen in a special way: 1) very close to the 
exact exponential solution (\ref{flat solution}), such
that the inflation lasts long enough. $\,$ 2) this choice 
must provide that, after
the inflationary phase ends, the Universe will approach 
the attractor corresponding to the FRW solution.
All the matter content 
of the Universe is created after the inflation ends through 
the decay of the massive degree of freedom induced by 
anomaly \cite{star,vile}. Unfortunately, despite 
this model is based on the QFT results
and does not need a special inflaton field, the amount 
of the fine-tuning for the initial conditions is at least
the same as for the inflaton-based models. 

In the recent works \cite{insusy,shocom,asta} we have 
developed an alternative version of the Starobinsky model, 
which does not require a fine-tuning for the initial data,
for it interpolates, naturally, between stable and unstable
regimes. In the rest of this article, we shall present a
brief exposition of our model.

\section{Inflation and SUSY, simple tests}

First of all, we rewrite the condition of stability 
$\,c>0\,$ in terms of the field content of the underlying QFT. 
We assume that the theory includes $N_0$ scalars, $N_{\frac12}$ 
fermions and $N_1$ vectors. The numbers 
$\,N_0,\,N_{\frac12},\,N_1\,$ reflect a particle 
content of the theory, and have nothing to do with
the real matter which might fill the Universe.  
Using a standard result (\ref{main equation}), we 
obtain \cite{wave}
\beq
c>0\quad\Longleftrightarrow\quad 
N_1 \,<\,\frac13\,N_{1/2}\,+\,\frac{1}{18}\,N_{0}\,.
\label{condition}
\eeq
The last inequality does not hold for the Standard Model, 
where $\,N_{1,\frac12,\,0}=(12,24,4)$. However, it 
is satisfied for its minimal supersymmetric extension 
(MSSM) with $\,N_{1,\frac12,\,0}=(12,32,104)$.
The same is true for any realistic supersymmetric
model, because the supersymmetrization procedure implies 
introducing fermion and scalar superpartners (sparticles) 
while the fundamental interactions (corresponding to the 
content of vector fields) are kept
the same. We can see that the transition between stable 
and unstable inflation can be associated with the SUSY 
breaking. Usually, the SUSY breaking implies the special 
form of the mass spectrum, when all sparticles are 
very heavy compared to observable particles (soft breaking). 
Therefore 
it is clear that inflation becomes unstable when its 
energy scale becomes less than the masses of the most 
of the sparticles and these sparticles decouple. 

All the time we will be concerned by the Feynman diagrams 
including the loops of the matter fields with external 
gravitational tails. For this reason, the typical energy 
for us is the energy of gravitons. In the cosmological 
setting, we shall associate it with the magnitude of the 
Hubble parameter $\,H$. 
Let us introduce the notation $\,M_*\,$ for the energy 
scale where the inequality (\ref{condition}) changes its 
sign to the opposite. The 
anomaly-induced inflation model assumes that the value of
$\,H\,$ is decreasing with time $\,{\dot H} < 0\,$ and that
$\,H_S\,$ is just an initial value of $\,H$. The stable 
inflation becomes unstable at the instant $\,t_f\,$ which 
is defined as a solution of the equation $\,H(t_f)=M_*$. 

It is worth mentioning two relevant results of QFT in curved 
space-time. 1) The decoupling of the loops of massive fields 
in curved space-time really takes place \cite{apco,fervi}. In 
particular, one can observe the smooth and monotone evolution 
of the coefficient $\,c\,$ with scale and also the change of 
its sign  from positive to negative due to the decoupling of 
the sparticles \cite{fervi}. 2) The coefficient $\,c\,$ 
in Eq. (\ref{main equation}) contains an arbitrariness
related, primarily, with the possibility to add 
$\,\int\sqrt{-g}R^2$-term to the classical action of vacuum.
However, this arbitrariness does not contradict our version
of the anomaly-induced inflation (see \cite{anom} for the 
details) because it can be always fixed by the 
renormalization condition. It is important that the 
ambiguity does not impose new constraints on the model 
and does not require a fine-tuning in the mentioned 
renormalization condition.

Before we proceed with inflation, let us 
perform two simple tests of the model.

\noindent
{\bf First test.} 
Consider the late FRW Universe with $k=\Lambda=0$:
The typical value of the Hubble parameter is 
$H_0\sim 10^{-42}\,GeV$. Then,
all massive particles decouple and the unique contribution 
to the anomaly comes from photon 
$\,N_{1,\frac12,\,0}=(1,0,0)$. It is easy to see that 
$c<0$ and the ``fast'' inflation is unstable.
Consider $\,a(t)\sim t^{2/3}$ in the equation (\ref{foe})
without CC and
inserting the dust-like source $\,{\rho_m}/(a^3\,c)\,$
in the {\it r.h.s}
\beq  
\frac{{\stackrel{....}{a}}}{a}
+\frac{{3\stackrel{.}{a}} {\stackrel{...}{a}}}{a^2}
+\frac{{\stackrel{..}{a}}^{2}}{a^{2}}
-\left( 5+\frac{4b}{c}\right) 
\frac{{\stackrel{..}{a}} {\stackrel{.}{a}}^{2}}{a^3}
-2k\left( 1+\frac{2b}{c}\right)
 \frac{{\stackrel{..}{a}}}{a^{3}}
=\frac{M_{P}^{2}}{8\pi c}
\left( \frac{{\stackrel{..}{a}}}{a}+
\frac{{\stackrel{.}{a}}^{2}}{a^{2}}\right)
-\frac{\rho_m}{c a^3}\,.
\label{matter}
\eeq
For the large values of time the ``classical''
(Einstein and the dust sourse) terms in the {\it r.h.s}
behave like $1/t^2$, while the fourth-derivative quantum 
corrections in the {\it l.h.s} behave as $1/t^4$.
Then, the quantum corrections are becoming irrelevant at
$t\to \infty$.

\noindent
{\bf Second test.} 
Consider the same physical situation as above 
but this time without
matter and with a small positive CC. In this case, according
to (\ref{HH}), $\,H=H_c=\sqrt{\Lambda/3}$. We 
will now check whether this solution is stable with respect
to perturbations of $H$. Consider
$H \to H_c + \mbox{const} \times e^{\la t}$.
The characteristic equation for $\la$ has the form
$$
\la^3+7H_c\la^2 + \left[\frac{(3c-b)4{H_c}^2 }{c}
- \frac{M^2_P}{8\pi c}\right]\la
- \,\frac{32\pi b{H_c}^3+M^2_PH_c}{2\pi c}\,=\,0\,.
$$
It is remarkable that all the solutions for the positive CC 
have negative real part
$$
\la_1=-4H_c
\,,\qquad
\la_{2/3}\,
=\,-\frac32\,H_c\,\pm\,\frac{M_P}{\sqrt{8\pi |c|}}\,i\,,
$$
while for the zero CC these solutions have zero 
real part and one would need to study the nonlinear
approximation.
Thus, $\La>0$ stabilizes our universe with respect to 
the dangerous quantum corrections.

%%%%%%%%%%%%%%%%%%%%%%%%%%%%%%%%%%%%%%%%%%%%%%%%%%%%%%%%%%%%%
\section{Vacuum effects of massive fields}

At that point we can conclude that our choice of the overall 
$\int\sqrt{-g}R^2$-term does not lead to problems in the IR
regime and can safely proceed in the study of the UV energies,
where we expect to meet natural inflation due to the quantum
effects of matter fields. The inflation starts in the stable 
phase because the particle content $(N_0,N_{1/2},N_1)$ is 
supersymmetric, and then becomes unstable due to the SUSY 
breaking and the decoupling of the sparticles. The above
story looks very appealing because it does not involve any 
sort of fine-tuning, links inflation with SUSY and also links 
the SUSY breaking with the graceful exit. However, until now 
there was an obvious loophole in this story. 
We are considering the inflation derived from anomaly-induced 
action, that is the action that results from the quantum 
effects of massless conformal fields. At the same time, in 
order to use the notion of decoupling one has to evaluate 
the effects of massive fields. Furthermore we expect that,
for some reason, the value of the Hubble parameter $H$ 
will decrease during inflation. But this would never happen 
if we have only massless fields, for in this case the 
inflation is exponential and $H=H_S$ is a constant. Hence,
our main hope is that taking masses of the fields into 
account we will really see the inflation slowing down.

In general, the problem of deriving the effective action
of vacuum for the massive fields is not solved yet. The 
existing regular methods, like covariant Schwinger-DeWitt
method, correspond to the expansion in the series in 
curvatures and their derivatives, divided by the 
corresponding powers of the particle masses. Therefore,
these methods are efficient only for the limit of large 
masses or, contrary to that, in the massless limit 
where the effective action can be obtained through
integrating the conformal anomaly.
In our case, the effective action of vacuum should be
calculated in the small-mass limit where the mentioned 
regular methods are not applicable. That is why the 
calculations Anzats for this case has been developed in 
\cite{shocom} and generalized
in \cite{asta}. Let us discuss it here in some details.

The idea is very simple: we formulate massive fields as
massless and conformal, by introducing a new scalar field 
$\chi$ and a new massive parameter $M$. The conformization 
of the Einstein-Hilbert action has been known for a time
\cite{Deser70} and for the massive matter fields it was
known at the level of a dilatation symmetry within the 
cosmon model \cite{PSW}. The 
conformal symmetry involving $\chi$ is reflected by the 
new Noether identity. This identity is anomalous, and 
integrating anomaly one arrives at the effective action.
After that the new degree of freedom $\chi$ is frozen 
and we arrive at the effective action of massive fields.

The first step is to introduce the conformal representation
of the massive fields. This can be achieved by replacing
\beq
m_{s,f} \to \frac{m_{s,f}}{M}\,\chi 
\,,\qquad
\frac{1}{16\pi G}\,R \,\to\, \frac{M_P^2}{16\pi M^2}\,
\left[\,R\chi^2 + 6\,(\pa \chi)^2\,\right]
\,,\qquad
\La \to \frac{\La}{M^2}\,\chi^2 \,,
\label{replace}
\eeq
where $m_{s,f}$ are scalar and fermion masses.

The divergences of the theory in the conformal 
representation have the form 
\beq
\Ga^{(1)}_{div} = \frac{\mu^{n-4}}{4-n}
\int d^nx\sqrt{|g|}\Big\{ wC^2 + bE + c{\nabla^2}R 
% \nonumber
% \\
+\frac{\tilde{f}M_P^2}{4\pi M^2}[R\chi^2 + 6(\pa \chi)^2]
+\frac{\tilde{g}M_P^4\chi^4}{4\pi M^4}\Big\},
\label{div}
\eeq
where 
\beq
\tilde{f} = 
\frac{1}{3\pi}\,\sum_{f}\,\frac{N_fm_f^2}{M_P^2}\,,
\qquad \mbox{and} \qquad
% \nonumber
% \\
\tilde{g} 
=\frac{1}{4\pi}\,\sum_{s}\frac{N_sm_s^4}{M_P^2\La}
\,-\,\frac{1}{\pi}\,\sum_{f}\frac{N_fm_f^4}{M_P^2\La}\,.
\label{replace11}
\eeq
Here the sums are taken over all species of fermions and
scalars with masses $\,m_f,\,m_s\,$ and multiplicities 
$\,N_f,\,N_s$. 

The classical Noether identity for the vacuum part of 
the effective action has the form
\beq 
{\cal T}\, = \,-\,\frac{2}{\sqrt{-g}} g_{\mu\nu}
\,\frac{\de S_{vac}}{\de g_{\mu\nu}}
+  \frac{1}{\sqrt{-g}}\,\chi\,
\frac{\de S_{vac}}{\de \chi}\,= \,0\,.
\label{vacu}
\eeq
The identity above differs from the usual conformal 
Noether identity $<T_\mu^\mu>=0$ due to the presence of 
$\,\chi$. Correspondingly, the conformal anomaly means 
$\,<{\cal T}>\,\neq\,0$ instead of usual 
$\,<{T}^\mu_\mu>\,\neq\,0$. Simple calculations give
\cite{shocom,asta}
\beq
<{\cal T}> \,=\, -\, \Big\{\,\,
wC^2\,+\,bE\,+\,c{\nabla^2}R 
\,+\,\frac{\tilde{f}\,M_P^2}{4\pi M^2}
\,[R\chi^2+6(\pa \chi)^2]
\,+\,\frac{\tilde{g}\,M_P^4}{4\pi M^4}\,\chi^4\,\,\Big\}\,.
\label{trace anomaly}
\eeq
After deriving the anomaly-induced effective action
and fixing the conformal unitary gauge
$\,\chi = {\bar \chi}\,e^{-\si}=M$, 
the one-loop effective action becomes
$$
{\bar \Gamma}^{(1)} = \int d^4 x\sqrt{|{\bar g}|} \big\{
w{\bar C}^2\sigma + b({\bar E} -\frac23 {\bar \nabla}^2 {\bar R})
\sigma+2 b\sigma{\bar \Delta}\sigma \}
- \frac{3c+2b}{36}\int d^4x\sqrt{-g}R^2
$$
\beq
-\int d^4 x\sqrt{|{\bar g}|} \Big\{
\frac{e^{2\si}}{16\pi G}[{\bar R}+6({\bar \na}\si)^2]
\big[1 - \tilde{f}\si\big]
- \frac{\La\, e^{4\si}}{8\pi G}\,
\big[1-\tilde{g}\si\big]\Big\} + S_c[g_{\mu\nu},M]\,,
\label{quantum for massive}
\eeq
where $\si=\ln a$ and $S_c[g_{\mu\nu},M]$ is the unknown
functional which is a constant of integration for the 
anomaly-induced effective action. In the case of massive
field this functional is not conformal invariant (it would 
be if we do not replace $\chi\to M$), and therefore the
formula above is just an approximation. The comparison with 
the renormalization group corrected classical action of 
vacuum shows that (\ref{quantum for massive}) is direct
generalization of it, with the usual scaling parameter 
$t$ (see, e.g. \cite{book})
substituted by the local quantity $\si$. Therefore, 
(\ref{quantum for massive}) must be a reliable approximation 
for the small-mass limit, exactly in the region where the 
Schwinger-DeWitt expansion is not efficient. 

According to the Eq. (\ref{quantum for massive}), the 
leading effect of the particle masses is that $1/G$ and 
the CC are replaced by the variable expressions
\beq
M^2_P\,\to\,M^2_P\,(1-\tilde{f}\ln a)\,,
\label{central G}
\\
\La\,M^2_P\,\to\,\La\,M^2_P\,(1 - \tilde{g}\ln a)\,.
\label{central sigma C}
\eeq
The last formulas show that, in principle, the 
effect of masses is slowly accumulating when the inflation
goes on. The reason is that, the dependence is logarithmic
and moreover the quantities (\ref{replace11}) are very 
small for any GUT, and incredibly small for, e.g. MSSM.

The equation for the conformal factor following from 
the action (\ref{quantum for massive}) has the form of 
(\ref{foe}) with the replacement (\ref{central G}),
plus some non-essential terms \cite{asta} with an 
additional factor of $\tilde{f}$. Therefore, we can 
expect that the procedure (\ref{central G}) should
also provide an approximate solution on the basis of the
exact stable inflationary solution (\ref{HH}). There is,
however, a strong constraint related to the value of 
$\tilde{g}$. At the beginning of inflation we assume
a small $\,\La\ll M_P^2$, that is why the solution
(\ref{HH}) does not depend on CC. But, when the 
inflation evolves, the Hubble parameter (\ref{HH}) will 
decrease according to (\ref{central G}) and the absolute 
value of the CC will increase very fast according to 
(\ref{central sigma C}). Then, instead of the graceful 
exit to the approximate FRW with the small value of the
CC, the universe will end up with the new phase of
inflation, driven by the $\,\La M^2_P\tilde{g}\si$
term. Indeed, the region when this terms becomes large,
is close to the limit of validity of the approximation
behind (\ref{quantum for massive}), but in order to 
perform the preliminary analysis it is better to impose
a constraint on the particle spectrum of the SUSY 
model and request that the $\be$-function for the CC
equals zero. Then, $\tilde{g}=0$ too, and the analysis
gets simplified.

The relation (\ref{central G}) can be easily rewritten
as a differential equation for the conformal factor
$\,H={\dot \si}=H_S\cdot (1-\tilde{f}\si)$, and the 
last can be solved immediately to give the following 
approximate analytical solution 
\beq
\si(t)\,=\,H_0\,t\,-\,\frac{H^2_0}{4}\,\tilde{f}\,t^2\,.
\label{parabola}
\eeq
It is interesting that the numerical analysis confirms
the parabolic dependence (\ref{parabola}) with enormous
precision \cite{asta}. 

The relation (\ref{parabola})
can be used to evaluate the total number of the inflationary 
$\,e$-folds for different models of the SUSY breaking. 
The first option is MSSM with the value $M_*\propto 1\,TeV$.
It is easy to see that in this case 
$\,\tilde{f}\propto (M_*/M_P)^2 = 10^{-32}$ and 
therefore the total amount of the $\,e$-folds is $10^{32}$.
The expected temperature of the Universe after the end
of inflation can be evaluated from Einstein equation 
in a usual way $\,T \propto \sqrt{M_*\,M_P} = 10^{11}\,GeV$,
which is a standard estimate for the inflaton-based models.
Alternatively, one may consider the SUSY breaking  
at the GUT scale. Suppose $M_*\propto 10^{14}\,GeV$.
Then the total amount of $\,e$-folds is about $10^{10}$
and the expected temperature after the end of inflation is  
high $\,T \propto 10^{16}\,GeV$. In this case the inflation 
does not solve the monopole problem of GUT's. Hence, the 
anomaly-induced inflation really favors low-energy SUSY.
Indeed, the intermediate versions (like, e.g. the Pati-Salam 
model) with the SUSY breaking at $10^{10}\,GeV$ are also 
possible. In this case we obtain $\,T \propto 10^{14}\,GeV$
which is better with respect to the monopole problem.

%%%%%%%%%%%%%%%%%%%%%%%%%%%%%%%%%%%%%%%%%%%%%%%%%%%%%%%%
\section{Problems of stability}

The stability of the inflationary solution from the 
initial stage until the graceful exit represents 
one more consistency test of the model. At the beginning the
role of the masses of the quantum fields is negligible and
the criterion of stability is given by (\ref{condition}).
Consider the later phase of inflation, when the quantum 
effects of massive fields temper exponential behavior. 
In this case we can use the approximate analytic
method and also numerical simulations. The results of both
methods are the same \cite{asta}. Let us briefly 
describe the analytic method. The stability or 
instability with respect to the small perturbations 
depends on the behavior of $\,\si(t)\,$ at the relatively 
small intervals of time,
where the Hubble parameter $\,H\,$ can be treated as a 
constant. Of course, when we move from one such interval 
to another, $\,H\,$ changes providing a source
for the perturbations. The direct calculations give the 
following equation for the perturbations 
$\,\si \to \si + y(\tau)$, where we used ``renormalized'' 
time variable $\,\tau=t/H$, $\,$ $H=const$:
\beq
{\stackrel{....}{y}} + 7\,{\stackrel{...}{y}}
+ 2\Big(6-\frac{b}{c}\Big)\,{\stackrel{..}{y}}
-\frac{8b}{c}\,{\stackrel{.}{y}} 
-\frac{4b}{c}\,\tilde{f}\,y = 0\,
\label{perturbation}
\eeq
At this point we
assumed, as before, a relatively small value of 
the cosmological constant. The last equation has a very 
special form, because all the coefficients are constants
and all but the last have the magnitude 
of the order one. The last coefficient is extremely 
small because of the factor $\,\tilde{f}\ll 10^{-9}$. The 
stability of equation with constant coefficients 
may be explored, e.g. using the Routh-Hurwitz (RH) 
conditions. A priory the RH determinants may have an 
arbitrary sign, but in our case they all turn out to be 
positive, such that the stability of the solution for 
tempered inflation (with respect to the perturbations of 
the variable $\,\sigma$) holds until the universe 
enters the marginal region $H\approx M_*$. 

%%%%%%%%%%%%%%%%%%%%%%%%%%%%%%%%%%%%%%%%%%%%%%%%%%%%%%%
%%%%%%%%%%%%%%%%%%%%%%%%%%%%%%%%%%%%%%%%%%%%%%%%%%%%%%%
%%%%%%%%%%%%%%%%%%%%%%%%%%%%%%%%%%%%%%%%%%%%%%%%%%%%%%%
\FIGURE[t]{\epsfig{file=oscilH.eps, width=.5\textwidth}
\caption{\sl Oscillations of $H(t)$ at the last stage of 
the stable inflation. Illustrative plot for
$\tilde{f}=10^{-5}$.}
\label{Figure 2}}	
%%%%%%%%%%%%%%%%%%%%%%%%%%%%%%%%%%%%%%%%%%%%%%%%%%%%%%%
%%%%%%%%%%%%%%%%%%%%%%%%%%%%%%%%%%%%%%%%%%%%%%%%%%%%%%%
%%%%%%%%%%%%%%%%%%%%%%%%%%%%%%%%%%%%%%%%%%%%%%%%%%%%%%%

The numerical analysis shows that, at the final stage of 
the stable anomaly-induced inflation, when the value of 
Hubble parameter is approaching $\,M_*$, this parameter
starts to oscillate (see Fig. 2). Indeed, these 
oscillations are due to the perturbations introduced by
the change of one ``constant'' value of $H$ to another.

The last testing of the model which has been performed 
so far is the stability with respect to tensor 
perturbations of the metric. In the covariant formalism 
(see, e.g., \cite{brandenberg}) the evolution of the tensor
degree of freedom is described by the coordinate-dependent 
scalar factor $\,h(t,{\vec r})\,$ of the tensor mode. 
The dynamical equation for $\,h(t)\,$ is very 
complicated \cite{star1,wave}, even for the 
theory of massless fields. Moreover, this equation
contains an ambiguity due to the conformal functional 
$\,S_c[g_{\mu\nu}]$ \cite{buodsh,wave}. One can partially
fix this ambiguity by choosing the proper vacuum for the 
perturbations \cite{wave}. As a result we meet an almost 
flat spectrum of the perturbations, however their 
amplitude may increase very fast. At the same time the
amplification of the amplitudes is performing much slower 
than the expansion of the conformal factor. As a result
the total metric becomes more and more homogeneous and
isotropic. This is a situation at the initial stage of 
inflation. 

At the last stage, e.g. in the last 65 $\,e$-folds, 
the equation for $\,h(t)\,$ is greatly simplified due
to the enormous number of the total $\,e$-folds between
the beginning and the end of inflation. The typical value 
for $\,\si\,$ depends on the model but, as we discussed
above, it varies between $10^{10}$ and $10^{32}$. In the 
last 65 inflationary $\,e$-folds the $\,\si\,$ itself may be 
treated as a big number. This feature greatly simplifies the 
equation for $\,h \equiv h(t,{\vec x})$, which can be 
presented as follows
$$
  b_0\stackrel{....}{h}
+ b_1\stackrel{...}{h}
+ b_2\stackrel{..}{h}
+ b_3\stackrel{.}{h}
+ b_4h+
$$
\beq
+n_1e^{-2\si}\na^2\stackrel{.}{h}
+n_2e^{-2\si}\na^2\stackrel{..}{h}
+n_3e^{-4\si}\na^4h\,=\,0\,,
\label{wave}
\eeq
where
\beq
b_0 = b_0(t) = a_1+w\cdot\si(t)\,,
\qquad
b_1 = 6H\,b_0+2wH\,,
\nonumber
\\
b_2 = 11H^2\,b_0+H^2(c-b/2+7w)\,,
\nonumber
\\
b_3 = 6H^3\,b_0+H^3(3c-3b/2+5w)\,,
\qquad
b_4 = -12H^4b\,,
\label{beta}
\eeq
\beq
n_1 = -2H\,b_0\,,
\qquad
n_2 = -2\,b_0\,,
\qquad
n_3 = \,b_0\,.
\label{n}
\eeq
It is worth noticing that in the general case, without 
the approximation of constant $H$ and without treating
$\,\si\,$ as a big number, the equation for $\,h \,$ is 
much more complicated \cite{star1,wave,hhr}. 
But in the physical situation of interest the equation can 
be simplified even further. The terms with space derivatives 
$\,\na h(t,{\vec r})\,$ are suppressed by the factors of 
$\,\exp (-2\si)\,$ and therefore are negligible. 
Furthermore, we can divide the equation by $\,\si\,$ 
and see that all the coefficients become constants
with accuracy of $\,1/\si$. In particular, in the last 
terms we can safely replace $\,1/\si$ by $1/\si_f=\tilde{f}$.
This value $\,\si_f=\tilde{f}^{-1}$ corresponds to 
the point of transition from stable to unstable inflation.
The most 
important difference is that, in the limit of large 
$\,\si\,$ we do not meet an arbitrariness related to 
the conformal functional $\,S_c[g_{\mu\nu}]$ and to 
the choice of the classical action of vacuum.
In fact, the equation for $\,h(t,{\vec r})$
is completely defined by the universal $\be$-functions
$\,w,b,c\,$ for the vacuum parameters. In particular,
the difference between the equations of \cite{star1}
and \cite{wave} (it is due to the distinct choices of 
$\,S_c[g_{\mu\nu}]$) disappears in this approximation.

The remaining equation for $\,h(t)\,$ has the form
\beq
\stackrel{....}{h} + 6\stackrel{...}{h}
+ 11\stackrel{..}{h} + 6\stackrel{.}{h}
\,-\, \frac{12b}{w\si_f}\,h\,=\,0\,.
\label{G wave}
\eeq

It is remarkable that the general structure of the 
Eq. (\ref{G wave}) is quite similar to the 
one of the Eq. (\ref{perturbation}) for the 
perturbations of $\,\si(t)$. But it is 
even more remarkable that the solution of 
(\ref{G wave}) does not have growing modes. 
One of the roots of the characteristic equation has
the magnitude 
of the order of $\,\tilde{f}\,$ and others of the 
order of one, but all of them have negative real parts.
For the physically reasonable choice of the initial 
data the amplitude of the perturbations almost remains 
constant. The significant amplification of the tensor
perturbations takes place only for the waves with the 
energies close to the Planck one, where all our 
semiclassical approach is not consistent. 

We conclude that the stability of the 
inflationary solution with respect to the perturbations
of the conformal factor and the tensor mode of the metric
holds from the initial stage (when the 
quantum fields may be approximately considered massless)
until the scale $M_*$, when most of the sparticles
decouple and the inflation becomes unstable\footnote{The
credibility of the approximation of $H=\mbox{const}\,$
for the tensor perturbations is not a trivial question,
but its verification requires significant new calculations
and therefore will be reported elsewhere.}. 
Indeed, the stability of the model may be jeopardized 
in the transition period, where we expect to 
meet rapid oscillations of the conformal factor which 
should lead to reheating. The main line in the 
further development of the model must be related 
with the quantitative model description of the decoupling
and the transition period.

%%%%%%%%%%%%%%%%%%%%%%%%%%%%%%%%%%%%%%%%%%%%%%%%%%%%%%%%%
\section{Concluding remarks}

We have briefly described the model of inflation based on 
the effective action of vacuum. This effective action follows
from  the quantum corrections of the massive fields. Indeed,
our approach to the effective action of the massive fields
is based on the special Anzats \cite{shocom,asta} which is 
reliable at the beginning of inflation, when masses are 
small compared to $H$. Due to the unbroken supersymmetry, 
at this scale the coefficient $\tilde{g}=0$ and the CC term 
is irrelevant. In this case the formula (\ref{parabola})
describes the evolution of the Universe, and the Hubble 
parameter is decreasing linearly with time. If we continue 
this evolution until the point $H=M_*$, the sparticles 
should decouple and the universe starts a new unstable 
phase of the evolution. We assume that this stage ends
in the FRW phase or in the present-day state with the 
small positive CC (\ref{HH}). However, it is easy to see
an inconsistency in the consideration presented above. In 
fact, our Anzats becomes not reliable when the value of 
the Hubble parameter $H$ is approaching $H=M_*$, because
this is the scale comparable to the masses of many 
sparticles. Similarly, at this scale the standard
curvature expansions (e.g. Schwinger-DeWitt) are also 
not applicable, because the masses of the particles 
and the Hubble parameter $H$ are of the same order of
magnitude in this region. All in all, what we have at the
moment is the description of the asymptotic regimes. It is
not clear, however, how to achieve the qualitative 
description of the region $H\propto M_*$, which is indeed 
the most interesting phenomenologically. If we get such
a description, this can open the way to the investigation 
of the reheating and density perturbations (which 
were actually considered in \cite{mukh} for the original
Starobinsky model). 

Finally, despite the anomaly-induced inflation is 
not as developed as inflaton models, it
represents an attractive alternative to them. In particular,
even at the present state of knowledge we have some obvious
advantages, such as the possibility to avoid a standard 
fine-tuning in the choice of initial data, a good 
chance to achieve a natural graceful exit and also to 
control the amplitude of the gravitational perturbations. 
Only further theoretical and phenomenological study of this 
model and comparison 
with experimental/observational data may eventually 
confirm or rule out this model of inflation.
\vskip 5mm

\noindent
{\large\it Acknowledgments.} 
Authors are grateful to J. C. Fabris for collaboration in the
development of the model and also to I. I. Kogan, V. N. Lukash
and especially to A. A. Starobinsky for fruitful discussions
A.P. is grateful to FAPESP for the {\it recem doutor} fellowship.
The work of J.S. has been supported in part by MECYT and FEDER
under project FPA2001-3598.
I.Sh. is grateful to the organizers of the workshop for 
partial financial support and to the CNPq for the fellowship.

%%%%%%%%%%%%%%%%%%%%%%%%%%%%%%%%%%%%%%%%%%%%%%%%%%%%%%%%%%%%%%%%%%%


\begin{thebibliography}{99}
%% \footnotesize

\bibitem{Guth} A.H. Guth, Phys. Rev. {\bf 23D} (1981) 347;
see also the review A.H. Guth, Phys. Rept. {\bf 333} (2000) 555.

\bibitem{linde} A. Linde, Phys. Lett. {\bf 108B} (1981) 389;

A. Albrecht and P.J. Steinhardt, 
Phys. Rev. Lett. {\bf 48} (1982) 1220.

\bibitem{KT} E. Kolb and M. Turner, {\sl The very early Universe}
                  (Addison-Wesley, New York, 1994);

E.Kolb, {\sl Cosmology and the origin of structures}.
Lectures at the X Brazilian School of Cosmology and Gravitation.
Mangaratiba, August, 2002.

\bibitem{fhh} M.V. Fischetti, J.B. Hartle and B.L. Hu, 
              Phys. Rev. {\bf D20} (1979) 1757.

\bibitem{star} A.A. Starobinsky, Phys. Lett. {\bf 91B} (1980) 99.

\bibitem{vile} A. Vilenkin, Phys. Rev. {\bf D32} (1985) 2511.

\bibitem{ander} P. Anderson, Phys. Rev. {\bf D28} (1983) 271;
{\bf D29} (1984) 615; {\bf D29} (1986) 1567.

\bibitem{birdav} N.D. Birrell and P.C.W. Davies, 
{\sl Quantum fields
in curved space} (Cambridge Univ. Press, Cambridge, 1982).

\bibitem{buodsh}
I.L. Buchbinder, S.D. Odintsov and I.L. Shapiro,
Phys. Lett. {\bf 162B} (1985) 92.

\bibitem{book} I.L. Buchbinder, S.D. Odintsov and I.L. Shapiro,
{\sl Effective Action in Quantum Gravity} (IOP Publishing,
Bristol, 1992).

\bibitem{asta} A.M. Pelinson, I.L. Shapiro and F.I. Takakura,
Nucl. Phys. {\bf 648B} (2003) 417.

\bibitem{SN} S. Perlmutter \textit{et al.}, {Astrophys. J.} 
{\bf 517} (1999) 565; 

A.G. Riess \textit{ et al.}, 
{Astrophys. J.} {\bf 116} \thinspace (1998) 1009.

\bibitem{wave} J.C. Fabris, A.M. Pelinson, I.L. Shapiro,
Grav. Cosmol. {\bf 6} (2000) 59; Nucl. Phys. {\bf B597} (2001) 539.

\bibitem{insusy} I.L. Shapiro, 
Int. Journ. Mod. Phys. {\bf 11D} (2002) 1159.

\bibitem{shocom} I.L. Shapiro, J. Sol\`{a},
Phys. Lett. {\bf 530B} (2002) 10.

\bibitem{apco} E.V. Gorbar and I.L. Shapiro,
JHEP {\bf 02} (2003) 021.

\bibitem{fervi} E.V. Gorbar and I.L. Shapiro,
JHEP {\bf 06} (2003) 004.   

\bibitem{anom} M. Asorey, E.V. Gorbar and I.L. Shapiro,
Class. Quant. Grav. {\bf 21} (2003) 163.   

\bibitem{Deser70} S. Deser, Ann. Phys. {\bf 59} (1970) 248.

\bibitem{PSW} R.D. Peccei, J. Sol\`{a}, C. Wetterich,
Phys. Lett. {\bf 195B} (1987) 183.

\bibitem{brandenberg} V.F. Mukhanov, H.A. Feldman 
and R.H. Brandenberger, Phys. Rep. {\bf 215} (1992) 203.

\bibitem{star1}
A.A. Starobinsky, JETP Lett. {\bf 30} (1979) 719; 
{\bf 34} (1981) 460.

\bibitem{hhr} S.W. Hawking, T. Hertog and H.S. Real,
Phys.Rev. {\bf D63} (2001) 083504.

\bibitem{mukh} 
V.F. Mukhanov and G.V. Chibisov, JETP Lett.
33 (1981) 532; JETP (1982) 258.

\end{thebibliography}
\end{document}